\documentclass[prd,floatfix,preprintnumbers]{revtex4}

\usepackage{graphicx}
\usepackage{graphicx,epsfig}
\usepackage{bm}
\usepackage{latexsym,amssymb,amsmath,float}

\newcommand {\ga} {\ {\raise-.5ex\hbox{$\buildrel>\over\sim$}}\ }
\newcommand {\la} {\ {\raise-.5ex\hbox{$\buildrel<\over\sim$}}\ }

\def\be{\begin{equation}}
\def\ee{\end{equation}}
\def\ba{\begin{eqnarray}}
\def\ea{\end{eqnarray}}

\begin{document}

\title{Exact general solutions for cosmological scalar field evolution in a background-dominated expansion}
\author{Robert J. Scherrer}
\affiliation{Department of Physics and Astronomy, Vanderbilt University,
Nashville, TN  ~~37235}

\begin{abstract}
We derive exact general solutions (as opposed to
attractor particular solutions) and corresponding first integrals for the
evolution of a scalar field $\phi$ in a universe
dominated by a background fluid with
equation of state parameter $w_B$.  In addition to the previously-examined
linear [$V(\phi) = V_0 \phi$] and
quadratic [$V(\phi) = V_0 \phi^2$] potentials, we
show that exact solutions exist
for the power law potential $V(\phi) = V_0 \phi^n$ with $n = 4(1+w_B)/(1-w_B) + 2$ and
$n = 2(1+w_B)/(1-w_B)$.  These
correspond to the potentials $V(\phi) = V_0 \phi^6$ and $V(\phi) = V_0 \phi^2$ for matter domination
and $V(\phi) = V_0 \phi^{10}$ and $V(\phi) = V_0 \phi^4$ for radiation domination.  The
$\phi^6$ and $\phi^{10}$ potentials can yield either oscillatory or non-oscillatory evolution,
and we use the first integrals to determine how the initial conditions map onto each form of evolution.
The exponential potential yields an exact solution for a stiff/kination ($w_B = 1$) background.  We use this exact
solution to derive an analytic expression for the evolution of the equation of state parameter, $w_\phi$,
for this case. 

\end{abstract}

\maketitle

\section{Introduction}

Scalar fields providing a significant component
of the energy density of the universe have frequently been invoked in cosmology.  They were first
introduced as the
main component of models for inflation (see, e.g., Refs. \cite{Lyth,Allahverdi} for reviews). Later, under the name ``quintessence," scalar
fields were introduced as an alternative to the cosmological constant as a mechanism to drive
the observed accelerated expansion
of the universe
\cite{RatraPeebles,Wetterich1,Wetterich2,Ferreira1,Ferreira2,CLW,CaldwellDaveSteinhardt,LiddleScherrer,SteinhardtWangZlatev}.
(See Refs. \cite{Copeland1,Bamba} for reviews).
More recently, the possibility that a scalar field might
contribute subdominantly to the expansion has been proposed as a possible solution
to the ``Hubble tension," the discrepancy between direct local measurements of the Hubble parameter
and the value inferred from measurements of the cosmic microwave background \cite{Poulin,Agrawal,Smith}.
(The viability of this proposal remains controversial; see, e.g., Refs.
\cite{Hill,Ivanov,Banerjee,Vagnozzi} for arguments against it, and Refs. \cite{Smith2,
Hill2,Poulin3} for
counterarguments).

The equation governing the evolution of a scalar field $\phi$ in a potential $V(\phi)$ is
\begin{equation}
\label{phievol}
\ddot{\phi} + 3H\dot{\phi} + \frac{dV}{d{\phi}} = 0,
\end{equation}
where the Hubble parameter H is given by
\begin{equation}
\label{Hdef}
H \equiv \frac{\dot{a}}{a} = \sqrt{\rho/3}.
\end{equation}
In this equation, $a$ is the scale factor, $\rho$ is the total density, and we take $8 \pi G = c = \hbar = 1$ throughout.

For most choices of $V(\phi)$, Eq. (\ref{phievol}) is nonlinear, and exact solutions cannot be derived.  On the other hand,
particular solutions can often be determined for specific potentials of interest.  These particular solutions lack arbitrary constants and so cannot be
fit to a given set of initial conditions on $\phi$.  However, these solutions can often be shown, under certain conditions,
to act as attractors, so that they describe the asymptotic behavior of $\phi$ for a wide range of initial conditions.
These attractor solutions have been exhaustively studied \cite{RatraPeebles,Wetterich2,Ferreira1,Ferreira2,CLW,LiddleScherrer,SteinhardtWangZlatev}.

Here we are concerned with a more difficult issue:  are there any known exact solutions or first integrals
for Eq. (\ref{phievol})?  In the context of inflation, the appropriate choice for $\rho$ in Eq. (\ref{Hdef}) is the energy
density of the scalar field itself:
\begin{equation}
\label{rhophi}
\rho_\phi = \frac{\dot{\phi}^2}{2} + V(\phi). 
\end{equation}
In this case, the Hamilton-Jacobi formulation \cite{SalopekBond}, which involves taking $H$ as the dependent variable
and $\phi$ as the independent variable, allows Eq. (\ref{phievol}) to be rewritten as a first-order equation:
\begin{equation}
\label{HJ}
\left(\frac{dH}{d\phi}\right)^2 - \frac{3}{2} H(\phi)^2 = - \frac{1}{2} V(\phi).
\end{equation}
Using Eq. (\ref{HJ}), it is straightforward to begin with a desired choice for $H(\phi)$ and derive a corresponding potential
$V(\phi)$; however, the reverse in not true.
Eq. (\ref{HJ}) remains nonlinear and does not, in general, yield exact solutions for arbitrary choices of $V(\phi)$.
However, there is one exception:  for the case where $V(\phi)$ is an exponential
potential, it is possible to derive exact (parametric) solutions
\cite{Russo,Elizalde,Andrianov,Piedipalumbo}.  (See also the approach in Ref. \cite{Harko}).

When analyzing quintessence models instead of inflation, the choice for $\rho$ in Eq. (\ref{Hdef}) becomes
more complicated.  Now we must include both the density of the scalar field as well as the density of
any additional background (radiation or nonrelativistic matter).  Denoting the latter by $\rho_B$, we have
\begin{equation}
\label{rhotot}
\rho = \rho_B + \frac{\dot{\phi}^2}{2} + V(\phi).
\end{equation}
In general, Eq. (\ref{phievol}) is intractable in this case, although approximate solutions have been derived
for certain conditions on the potential when $\rho_B$ represents nonrelativistic matter \cite{ScherrerSen,ds1,Chiba,ds2}.
An exact solution was claimed for the case of nonrelativistic matter plus a scalar field with an exponential
potential in Ref. \cite{Basilakos}, but this solution was later shown to be flawed \cite{ACK}.

However, in some physically-interesting cases, the contribution of the scalar field to the energy density can be neglected
in comparison to the density of the background component.  It is these models
which are the subject of this paper.  For quintessence, this will be the case whenever
the quintessence density is initially subdominant and becomes important only at late times.  (This is effectively
the case for which attractor solutions are derived in Refs.
\cite{RatraPeebles,Wetterich2,Ferreira1,Ferreira2,CLW,LiddleScherrer,SteinhardtWangZlatev}).  Furthermore, the scalar field models
invoked to resolve the Hubble tension \cite{Poulin,Agrawal,Smith} assume a transient density contribution from the scalar
field that is always subdominant relative to the background (radiation or matter) density.

While a handful of exact solutions have previously been discussed for constant, linear, and quadratic potentials, we show here
that there are exact solutions for a variety of other potentials.  While almost none of these solutions are ``new" in the sense of
being unknown in the mathematics literature, they have not been previously applied to cosmological scalar field
evolution.  In the next section, we review the general evolution for a scalar field in a barotropic background.
In Sec. III, we re-examine the previously-derived exact solutions for constant, linear, and quadratic potentials.
In Sec. IV, we derive exact solutions and first integrals for power-law potentials, and we use the first integrals to
solve an interesting question for potentials that support both oscillatory and non-oscillatory behavior.  In Sec. V,
we examine exact solutions for exponential potentials.  We discuss our results briefly in Sec. VI.

\section{Scalar field equation of motion in a barotropic background}

We will assume that the expansion of the universe is dominated by a barotropic fluid with an equation of state parameter
$w_B \equiv p_B/\rho_B$, where $p_B$ and $\rho_B$ are the fluid pressure and density, respectively.  The
most important cases are radiation, with $w_B= 1/3$ and matter, with $w_B = 0$.  However, we will consider
arbitrary $w_B$; as we will see below, the case of stiff matter ($w_B = 1$) provides some particularly interesting results.
Then for $-1 < w_B \le 1$, the background density evolves as
\begin{equation}
\rho_B \propto a^{-3(1+w_B)},
\end{equation}
and the Hubble parameter is given by
\begin{equation}
H = \frac{2}{3(1+w_B)} \frac{1}{t},
\end{equation}
so that the evolution equation for $\phi$ in a background-dominated expansion is
\begin{equation}
\label{phiback}
\ddot{\phi} + \frac{2}{1+w_B}\frac{1}{t}\dot{\phi} + \frac{dV}{d{\phi}} = 0.
\end{equation}
This is the equation for which we seek exact solutions.
These exact solutions will yield two free parameters, which are determined by specifying the values of
$\phi$ and $\dot\phi$ at some fiducial initial time $t_i$; we will denote these by $\phi_i$ and $\dot \phi_i$,
respectively.

The physically-observable quantity is not the value of $\phi$, but the density $\rho_\phi$ given by
Eq. (\ref{rhophi}).  It is conventional to parametrize the evolution of $\phi$ in terms of the equation
of state parameter $w_\phi \equiv p_\phi/\rho_\phi$, where the scalar field pressure is given by
\begin{equation}
p_\phi = \frac{\dot{\phi}^2}{2} - V(\phi).
\end{equation}
Note that the evolution of $\phi$ is unaffected by addition of a constant to the density in Eq.
(\ref{phiback}).  However, this will affect $\rho_\phi$ and $w_\phi$, so we consider this possibility in discussing
our solutions below.

\section{Constant, linear, and quadratic potentials}

Here we examine potentials of the form
\begin{equation}
\label{linear}
V(\phi) = V_0 + V_1 \phi + V_2 \phi^2,
\end{equation}
where $V_0$, $V_1$, and $V_2$ are constants.  Background-dominated solutions for constant, linear, and quadratic
potentials have all been derived previously, but we include them here for completeness, and we consider
in addition linear
combinations of such potentials as in Eq. (\ref{linear}).  In all of these cases, Eq. (\ref{phiback}) reduces to a linear
differential equation with straightforward solutions.

Consider first the case of a constant potential $V(\phi) = V_0$.  Then $dV/d\phi = 0$, and Eq. (\ref{phievol})
can be solved for $\dot \phi(a)$ for arbitrary $H$ (not just the background-dominated case).  We obtain
$\dot \phi(a) = C_1 a^{-3}$,
where $C_1$ is a constant of integration.  Then the density as a function of the scale factor is just
\begin{equation}
\rho_\phi(a) = (C_1^2/2) a^{-6} + V_0,
\end{equation}
i.e., the density evolves as the sum of a constant-density component and a stiff-matter component.  Models of
this sort were dubbed ``skating" models and investigated previously in Refs. \cite{Linder,Sahlen}.

Now consider the linear potential $V(\phi) = V_1 \phi$.  For the background-dominated case, the evolution of
$\phi$ was first derived in Ref. \cite{LiddleScherrer}; in this case we have
\begin{equation}
\phi = C_1 + C_2 t^{(w_B - 1)/(w_B + 1)} - \left( \frac{1+w_B}{6+2w_B}\right)
V_1 t^2,
\end{equation}
for $w_B \ne 1$.  For the case of a stiff matter background ($w_B = 1$) we
have instead
\begin{equation}
\phi = C_1 + C_2 \ln(t) - \frac{1}{4}
V_1 t^2.
\end{equation}

Now consider the quadratic potential $V(\phi) = V_2 \phi^2$.  In this case,
Eq. (\ref{phiback}) becomes
\begin{equation}
\label{phiquad}
\ddot{\phi} +\frac{2}{1+w_B} \frac{\dot{\phi}}{t} + 2 V_2 \phi = 0,
\end{equation}
As noted in Ref. \cite{LiddleScherrer}, this is just a form of Bessel's
equation,
with general solution
\begin{equation}
\label{Bessel1}
\phi = t^{-\alpha}[C_1 J_\alpha(\sqrt{2V_2}t)+ C_2 Y_\alpha(\sqrt{2V_2}t)],
\end{equation}
where $J_\alpha$ and $Y_\alpha$ are Bessel functions of the first and second kind, and
$\alpha$ is given by 
\begin{equation}
\label{phiBessel}
\alpha =  \frac{1}{1+w_B} -\frac{1}{2}.
\end{equation}
When $\alpha$ is not an integer, the solution can be written in the simpler form
\begin{equation}
\label{Bessel2}
\phi = t^{-\alpha}[C_1 J_\alpha(\sqrt{2V_2}t)+ C_2 J_{-\alpha}(\sqrt{2V_2}t)],
\end{equation}
For the matter-dominated case, this solution takes the particularly simple form \cite{masso}
\begin{equation}
\label{Bessel3}
\phi = \frac{C_1 \sin(\sqrt{2V_2}t) + C_2 \cos (\sqrt{2V_2}t)}{t}.
\end{equation}

Finally consider the full potential given by Eq. (\ref{linear}).  In this case, the solutions
given by Eqs. (\ref{Bessel1}), (\ref{Bessel2}), and (\ref{Bessel3})
are simply modified by the addition of a constant:
\begin{equation}
\phi = t^{-\alpha}[C_1 J_\alpha(\sqrt{2V_2}t)+ C_2 Y_\alpha(\sqrt{2V_2}t)] - \frac{V_1}{2V_2},
\end{equation}
\begin{equation}
\phi = t^{-\alpha}[C_1 J_\alpha(\sqrt{2V_2}t)+ C_2 J_{-\alpha}(\sqrt{2V_2}t)]  - \frac{V_1}{2V_2},
\end{equation}
and
\begin{equation}
\phi = \frac{C_1 \sin(\sqrt{2V_2}t) + C_2 \cos (\sqrt{V_2}t)}{t} - \frac{V_1}{2V_2}.
\end{equation}
As an example, consider evolution in a matter-dominated background with the boundary conditions
$\phi = \phi_i$ and $\dot \phi =0$ in the limit $t \rightarrow 0$.  Then Eq. (\ref{Bessel3}) yields
\begin{equation}
\phi = \left(\phi_i + \frac{V_1}{2V_2}\right) \frac{\sin(\sqrt{2V_2}t)}
{\sqrt{2V_2}t}-\frac{V_1}{2V_2}.
\end{equation}
	
\section{Power-law potentials}

\subsection{Particular solutions}
Consider the evolution of a scalar field in a background-dominated expansion with a power-law potential of the form
\begin{equation}
V = V_0 \phi^n,
\end{equation}
with $V_0 > 0$.
Unlike the linear and quadratic potentials of the previous section, the exact solutions we will derive
shortly have not been previously discussed in
the context of cosmological scalar fields.  For these power-law potentials,
Eq. (\ref{phiback}) becomes
\begin{equation}
\label{phipower}
\ddot{\phi} + \frac{2}{1+w_B}\frac{1}{t}\dot{\phi} + nV_0 \phi^{n-1} = 0.
\end{equation}
Note that if $\phi(t)$ is a solution to Eq. (\ref{phipower}), then so is $C^{2/(n-2)}\phi(Ct)$,
where $C$ is an arbitrary constant.
Thus, any general solution to this equation
must be of the form
\begin{equation}
\label{generalpower}
\phi = C^{2/(n-2)}f(Ct),
\end{equation}
where $C$ accounts for one of the two arbitrary constants that must occur in the solution.

Power-law potentials in a background-dominated universe were first studied by Ratra and Peebles \cite{RatraPeebles} and
further examined in Refs. \cite{LiddleScherrer,SteinhardtWangZlatev}.  Eq. (\ref{phipower}) has a well-known particular
solution, namely
\begin{equation}
\label{particular}
\phi = \left[- \frac{1}{nV_0}\left(\frac{2}{2-n}\right)\left(\frac{n}{2-n} + \frac{2}{1+w_B}\right)\right]^{1/(n-2)} t^{2/(2-n)}.
\end{equation}
for $n<0$ or $n>2$.

Because Eq. (\ref{particular}) gives only a particular solution, there is no guarantee that this solution is
an attractor of the equation of motion, and when it is an attractor, there is no way to determine the evolution of $\phi$ as
this attractor is achieved.  Ratra and Peebles \cite{RatraPeebles} showed that this particular solution is, in fact,
an attractor for negative $n$ in a radiation or matter-dominated background.  Liddle and Scherrer \cite{LiddleScherrer}
extended this result to all values of $w_B$ for $n < 0$.

However, for $n>2$ the result is more complicated.  First, the particular solution for $n>2$ is valid only for
\begin{equation}
\label{condition}
n > \frac{4}{1-w_B}.
\end{equation}
Second, it can be shown that this solution is an attractor only for \cite{LiddleScherrer}
\begin{equation}
\label{attractor}
n > \frac{2(3+w_B)}{1-w_B}.
\end{equation}
Note that Eq. (\ref{condition}) is satisfied whenever Eq. (\ref{attractor}) is true.  The two cases of greatest
interest are radiation domination, for which $w_B = 1/3$ and the condition for the particular solution to be an
attractor is $n > 10$, and matter domination, for which $w_B = 0$ and the attractor condition gives $n > 6$.  When
the condition given by
Eq. (\ref{attractor}) is satisfied, $\phi$ evolves smoothly to zero as given by Eq. (\ref{particular}),
while when the particular solution is not an attractor, the solution is oscillatory about $\phi = 0$ (assuming
$n$ is an even integer or $\phi$ is replaced by $|\phi|$ in Eq. \ref{phipower}).
See Refs. \cite{Agrawal,ds3} for a further discussion
of these points. 
When $n = 2(3+w_B)/(1-w_B)$, the solution is neutrally stable, allowing both oscillating and non-oscillating trajectories
depending on the initial conditions \cite{LiddleScherrer}.  We will see that it is precisely these cases that yield
two of our exact solutions below.

\subsection{Exact solutions}

To search for exact solutions to Eq. (\ref{phipower}), we make the change of variables
\begin{equation}
t = f(\tau),
\end{equation}
and
\begin{equation}
\phi(\tau) = g(\tau)\psi(\tau),
\end{equation}
where the functions $f(\tau)$ and $g(\tau)$ will be chosen to produce an
exactly-solvable differential equation for $\psi(\tau)$.
With these substitutions, Eq. (\ref{phipower}) becomes
\begin{equation}
\psi^{\prime \prime} + \left[2 \frac{g^{\prime}}{g} + \frac{2}{1+w_B}
\frac{f^\prime}{f} - \frac{f^{\prime \prime}}{f^{\prime}}\right] \psi^\prime
+ \left[ \frac{g^{\prime \prime}}{g} + \frac{2}{1+w_B}
\frac{f^\prime}{f}\frac{g^\prime}{g} - \frac{f^{\prime \prime}}{f^{\prime}}
\frac{g^\prime}{g}\right]\psi + n V_0 f^{\prime 2} g^{n-2}\psi^{n-1} = 0,
\end{equation}
where the prime denotes derivative with respect to the new independent variable
$\tau$.  In order to derive a first integral, and thence an exact solution, we seek
functions $f(\tau)$ and $g(\tau)$ for which the factor multiplying $\psi^{\prime}$ is
zero, and the factors multiplying $\psi$ and $\psi^{n-1}$ are constants.
For each value of $w_B$, there are two values of $n$ that allow us to derive
such functions. We will refer to these as Case (A) and Case (B).

\vspace{0.3 cm}

\noindent Case A:
\begin{eqnarray}
n &=& 4 \left(\frac{1+w_B}{1-w_B}\right) + 2,\\
f(\tau) &=& \exp(\tau),\\
g(\tau) &=& \exp\left(\frac{2}{2-n}\tau\right).
\end{eqnarray}

\noindent Case B:
\begin{eqnarray}
n &=& 2\left(\frac{1+w_B}{1-w_B}\right),\\
f(\tau) &=& \tau^{n/2},\\
g(\tau) &=& 1/\tau.
\end{eqnarray}

For matter ($w_B = 0$) and radiation ($w_B = 1/3$)
dominated expansions, Case A corresponds to $n = 6$ and $n=10$, respectively.
As noted above, these are precisely the cases demarcating the boundary between
the oscillating and attractor (non-oscillating) solutions.  Case B corresponds
to $n=2$ for matter domination and $n=4$ for radiation
domination.  The former was already discussed in Sec. III, while the latter
will yield a new exact solution.

First consider Case A.  Under the transformation above, we obtain
\begin{equation}
\psi^{\prime\prime} - \frac{4}{(n-2)^2}\psi + nV_0\psi^{n-1} = 0,
\end{equation}
which can be integrated to yield
\begin{equation}
\label{intApsi}
\frac{1}{2} \psi^{\prime 2} - \frac{2}{(n-2)^2}\psi^2 + V_0 \psi^n = C,
\end{equation}
where $C$ is a constant determined by the boundary conditions.
In terms of the original variables $\phi$ and $t$, Eq. (\ref{intApsi}) is
\begin{equation}
\label{intA}
\frac{1}{2} t^{2n/(n-2)}\dot \phi^2 + \frac{2}{n-2}t^{(n+2)/(n-2)}
\phi \dot \phi + t^{2n/(n-2)} V_0 \phi^n = C.
\end{equation}
Eq. (\ref{intA}) provides a first integral for the evolution of the scalar field
for Case A.

We can now use Eq. (\ref{intApsi}) to derive an exact solution.  This equation
can be rewritten as
\begin{equation}
\tau = \int \frac{d\psi}{\sqrt{4 \psi^2/(n-2)^2 - 2 V_0\psi^n + C_1}} + \ln C_2
\end{equation}
The relation between $\tau$, $\psi$, $t$, and $\phi$ then gives
us an exact solution in parametric form:
\begin{eqnarray}
t &=& C_2\exp\left (\int\frac{d\psi}{\sqrt{4 \psi^2/(n-2)^2 - 2 V_0\psi^n + C_1}}\right),\\
\phi &=& C_2^{2/(2-n)}\psi\exp\left(\frac{2}{2-n} \int\frac{d\psi}{\sqrt{4 \psi^2/(n-2)^2 - 2 V_0\psi^n + C_1}}\right).
\end{eqnarray}
Note that this solution has the form of the general solution in Eq. (\ref{generalpower}),
with $C_2$ corresponding to $C$ in that equation.
As expected, our exact solution contains
two arbitrary constants, which are determined by the initial conditions
on $\phi(t)$ and $\dot \phi(t)$.

Now consider Case B.  The indicated transformation gives
\begin{equation}
\label{CaseBdiffeq}
\psi^{\prime \prime} + \frac{n^3}{4} V_0 \psi^{n-1} = 0,
\end{equation}
which integrates to
\begin{equation}
\label{intBpsi}
\frac{1}{2} \psi^{\prime 2} + \frac{n^2}{4}V_0 \psi^n = C,
\end{equation}
with $C$ a constant.  Reexpressing this equation in terms of $\phi$ and $t$
leads to the corresponding first integral for case B:
\begin{equation}
\label{intB}
\frac{1}{2} t^2 \dot \phi^2 + \frac{2}{n} t\phi \dot \phi + \frac{2}{n^2}\phi^2 + 
 t^2 V_0 \phi^n = C.
\end{equation}
Again, we can use Eq. (\ref{intBpsi}) to derive an exact parametric solution:
\begin{eqnarray}
t &=& C_2 \left(\int\frac{d\theta}{\sqrt{1-\theta^n}} + C_1\right)^{n/2}
,\\
\phi &=& C_2^{2/(2-n)} (n^2 V_0/2)^{1/(2-n)} \theta
\left(\int\frac{d\theta}{\sqrt{1-\theta^n}} +
C_1\right)^{-1},
\end{eqnarray}
where our new parametric variable is $\theta = C^{-1/n} \psi$.  As in the
previous case, our solution take the general form given by Eq.
(\ref{generalpower}).
Note that both sets of first integrals and exact solutions can be found, in a
slightly different form, in Ref. \cite{Polyanin}.

\subsection{Applications}

Consider first the set of solutions given by Case A.  As noted, these
correspond to $n=10$ for evolution in a radiation-dominated background
and $n=6$ in a matter-dominated background.  The latter is the more
interesting case, since it is close to the value of
$n$ required in models that use the additional
energy density from an evolving scalar field to
resolve the Hubble tension \cite{Poulin,Agrawal,Smith}.  Further, a scalar field
initially at rest in this potential could serve as a ``thawing" model for
quintessence.

Note that Eq. (\ref{phipower}) with $w_B = 0$ and $nV_0 = 1$ corresponds
exactly to the Lane-Emden equation, with $n=6$ resulting in a set of
previously-investigated particular solutions.  In addition to the particular
solution corresponding to Eq. (\ref{particular}), there is a classic
solution of the form
\begin{equation}
\label{classic}
\phi = \frac{1}{\sqrt{1+t^2/3}}.
\end{equation}
(See, e.g., Ref. \cite{Binney} for a detailed discussion of this solution).
Later, the solution
\begin{equation}
\phi = \frac{\sin(\ln \sqrt{t})}{\sqrt{3t - 2t \sin^2(\ln\sqrt{t})}}
\end{equation}
was discovered independently by Srivastava \cite{Srivastava} and Sharma \cite{Sharma}.
All of these solutions can be extended to a one-parameter family of solutions
using Eq. (\ref{generalpower}).  Finally, the full general solution
was derived by Mach \cite{Mach}, who showed that particular discrete
values of $C$ correspond to the previously-discovered exact solutions.
For example, $C=0$ gives the solution in Eq. (\ref{classic}).  Mach's
solutions for other values of $C$ are given in terms of Jacobi and Weierstrass elliptic
functions.

Because our own exact solution is in integral parametric form,
it is difficult to use, and
the first integral (Eq. \ref{intA}) actually provides more useful information.
For $n=6$, we obtain
\begin{equation}
\label{int6}
\frac{1}{2} t^{3}\dot \phi^2 + \frac{1}{2}t^{2}
\phi \dot \phi + t^{3} V_0 \phi^6 = C.
\end{equation}
This first integral is given by Leach \cite{Leach} who cites several
earlier references to it.  One of the more interesting aspects of scalar
field evolution in this case is the fact (emphasized by Refs.
\cite{LiddleScherrer,Agrawal,ds3}) that this potential can support both oscillatory
behavior as well as smooth evolution to $\phi=0$.  One might naively assume that a sufficiently
small initial value of $\dot \phi$ would allow
$\phi$ to evolve smoothly to zero, while a large negative value of $\dot \phi$ would always
lead to oscillatory behavior.  Suprisingly, the opposite is the case.

Eq. (\ref{int6}) provides
a simple condition to determine which form of evolution takes place for
a given set of initial conditions.  For simplicity, we will take the initial time to
correspond to $t=1$, with initial values of $\phi = \phi_i$ and
$\dot \phi = \dot \phi_i$.  Then it is easy to see
that the condition for oscillatory behavior
is $C > 0$, corresponding to
\begin{equation}
\frac{1}{2} \dot \phi_i^2 + \frac{1}{2}
\phi_i \dot \phi_i +  V_0 \phi_i^6 > 0.
\end{equation}
Thus, for a scalar field initially at rest ($\dot \phi_i=0$), all solutions
produce oscillatory behavior.  The only way for $\phi$ to evolve
smoothly to zero is for $\phi_i$ and $\dot \phi_i$ to
have opposite signs; i.e., the field must initially be rolling
downhill!  In order to achieve $C \le 0$, the
initial values for $\phi$ and $\dot \phi$ must satisfy
\begin{equation}
\phi_i < (8V_0)^{-1/4},
\end{equation}
and
\begin{equation}
\frac{\phi_i}{2}[-1-\sqrt{1-8V_0 \phi_i^4}] < \dot \phi_i <
\frac{\phi_i}{2}[-1+\sqrt{1-8V_0 \phi_i^4}],
\end{equation}
where we have taken $\phi_i > 0$.
When these conditions are satisfied, the field does not oscillate but instead evolves smoothly to zero.
The regions in parameter space for which these two types of behavior occur are illustrated in Fig.
1 for the case $V_0 = 1$.
\begin{figure}[t]
\centerline{\epsfxsize=3.7truein\epsfbox{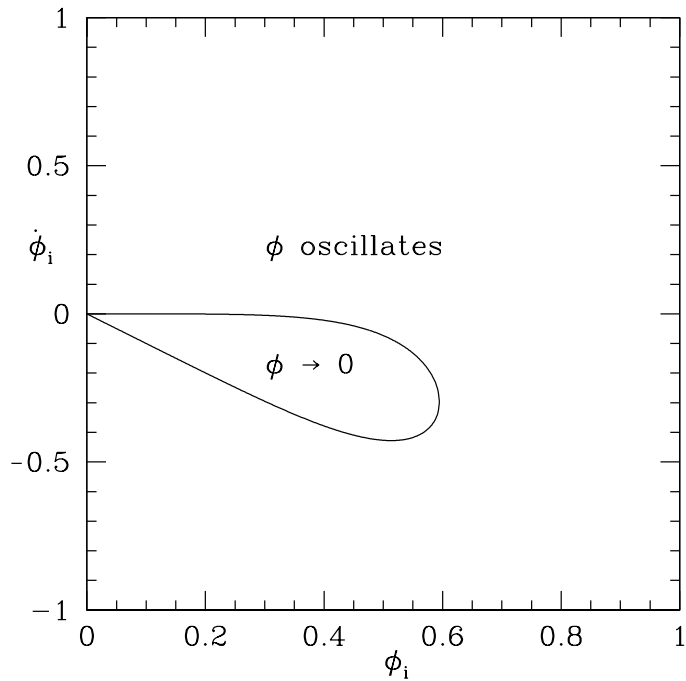}}
\caption{The regions in parameter space corresponding to oscillatory evolution of $\phi$
versus $\phi$ evolving smoothly to zero in a matter-dominated
background with the potential
$V = V_0 \phi^6$ with $V_0 = 1$
and the indicated initial
values of $\phi$ and $\dot \phi$ at $t_i = 1$.}
\end{figure}
\begin{figure}[t]
\centerline{\epsfxsize=3.7truein\epsfbox{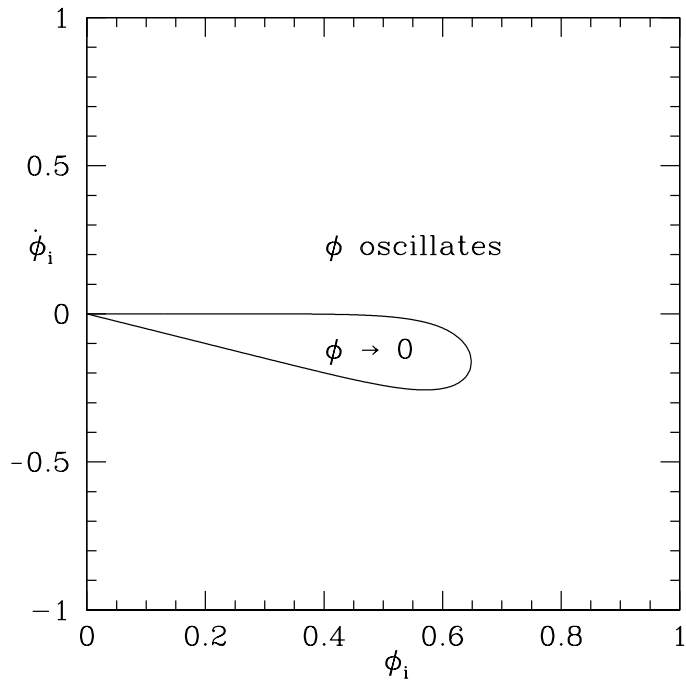}}
\caption{As Fig. 1, for
$\phi$ evolving in a radiation-dominated
background with the potential
$V = V_0 \phi^{10}$ with $V_0 = 1$
and the indicated initial
values of $\phi$ and $\dot \phi$ at $t_i = 1$.}
\end{figure}
Fig. 1 suggests that oscillatory behavior for the $\phi^6$ potential
during matter domination is in some sense
more ``generic" than $\phi \rightarrow 0$ evolution.
Although both types of
evolution are possible, the latter corresponds to a special
set of initial conditions with a finite range in both
$\dot \phi_i$ and $\phi_i$.

The other Case A solution of physical interest is a radiation-dominated
background with $n=10$.  In this case Eq. (\ref{intA}) gives the first integral
\begin{equation}
\frac{1}{2} t^{5/2}\dot\phi^2 + \frac{1}{4} t^{3/2} \phi \dot \phi
+ t^{5/2}V_0 \phi^{10} = C.
\end{equation}
The analysis here is qualitatively similar to the $n=6$ case for matter
domination.  This potential supports both oscillatory behavior and
smooth $\phi \rightarrow 0$ evolution, but the latter requires $\dot \phi_i <0$ for $\phi_i >0$ and, for an initial time $t=1$,
\begin{equation}
\phi_i < (32V_0)^{-1/8},
\end{equation}
and
\begin{equation}
\frac{\phi_i}{4}[-1-\sqrt{1-32V_0 \phi_i^8}] < \dot \phi_i <
\frac{\phi_i}{4}[-1+\sqrt{1-32V_0 \phi_i^8}].
\end{equation}
The regions in parameter space corresponding to these two types of behavior
are illustrated in Fig.
2 for the case $V_0 = 1$.
As in the case of matter domination, the region in which
the field evolves smoothly to zero represents a small fraction
of parameter space, suggesting that oscillatory behavior in this
case is the more generic behavior. 

Now consider Case B.  
Note that the first integral for Case B can be rewritten in the form
\begin{equation}
\frac{1}{2}\left(t \dot \phi + \frac{2}{n}\phi \right)^2 + t^2 V_0 \phi^n = C.
\end{equation}
For even $n$, we have $C > 0$, so these solutions always oscillate.

For a matter-dominated background, Case B corresponds to $n=2$,
which gives the trigonometric exact solution already discussed in Sec. III.  For
a radiation-dominated background, Case B corresponds to a quartic potential
($n=4$).
For $n=4$, Eq. (\ref{CaseBdiffeq}) can be solved to
express $\psi(\tau)$ in terms
of a Jacobi elliptic function; the resulting expression for $\phi(t)$ is
\begin{equation}
\label{Jacobi}
\phi = \frac{C_1}{\sqrt{t}}{\rm cn}[4C_1\sqrt{V_0}(\sqrt{t} - C_2),1/\sqrt{2}].
\end{equation}
where $C_1$ and $C_2$ are constants of integration.
This solution was derived by Greene et al. \cite{Greene} for evolution in a scalar-field
dominated background in the context of inflation.  The reason that Eq. (\ref{Jacobi}) gives the same answer for
a radiation-dominated background is that Greene et al. assumed a rapidly oscillating ($\nu \gg H$)
scalar field.
In this limit, with
a potential
of the form $V(\phi) = V_0 \phi^4$, the scalar field energy density driving 
the expansion decays as \cite{Turner}
$\rho_\phi \propto a^{-4}$.
Masso et al. \cite{masso} give an approximate analytic
solution in terms of Jacobi elliptic functions for the evolution of $\phi(t)$ in the $\phi^4$ potential for arbitrary $w_B$.
It is clear from our derivation
that their solution is exact for $w_B = 1/3$ and approximate for all other cases.

The first integral for $n=4$ is
\begin{equation}
\label{intn=4}
\frac{1}{2} t^2 \dot \phi^2 + \frac{1}{2} t\phi \dot \phi + \frac{1}{8}\phi^2 + 
 t^2 V_0 \phi^4 = C.
\end{equation}
The limit for which the oscillations are rapid in comparison to the expansion rate corresponds to
$t \gg 1$, so that the first and last terms on the left-hand side of Eq. (\ref{intn=4}) dominate,
giving $t^2 \rho_\phi = C$, which corresponds to $\rho_\phi \propto a^{-4}$ in a radiation-dominated background,
as expected.  How does $\rho_\phi$ evolve when the oscillation rate is not large compared to the expansion rate?  In
that case, we can use Eq. (\ref{Jacobi}) to calculate directly the deviation from $a^{-4}$ evolution,
but it is easier to use the first integral to gain some insight into this evolution.
If we consider the value of $\rho_\phi$ when the field is at the minimum
in the potential ($\phi = 0$), then we once again have $t^2 \rho_\phi = C$, i.e., the value of
the kinetic component of the energy when $\phi$ is at the minimum of the potential scales exactly like $a^{-4}$ even when the oscillation frequency
is not large compared to $H$.  In contrast, if we examine $\rho_\phi$ when the field achieves its maximum value, $\phi =
\phi_m$,
at
$\dot \phi = 0$, we have $t^2 \rho_\phi = C - (1/8) \phi_m^2$.  Thus $t^2 \rho_\phi$ is always smaller than its asymptotic value,
but this difference decays away as $\phi_m$ decreases with each oscillation.
\section{Exponential potentials}

\subsection{Particular solutions}

Now we assume a scalar field in a background-dominated expansion with an
exponential potential of the form
\begin{equation}
V = V_0 e^{-\lambda \phi},
\end{equation}
with $V_0 > 0$ and $\lambda > 0$.  Potentials of this form
were among the first to be examined as possible models for quintessence
\cite{Wetterich1,RatraPeebles,
Ferreira1,Ferreira2,CLW,LiddleScherrer}.
For these exponential potentials, Eq. (\ref{phiback}) gives
\begin{equation}
\label{phiexp}
\ddot{\phi} + \frac{2}{1+w_B}\frac{1}{t}\dot{\phi} - \lambda V_0 e^{-\lambda \phi}
 = 0.
\end{equation}
This has a well-known particular solution
\begin{equation}
\label{exppart}
\phi = \frac{2}{\lambda} \ln t + \frac{1}{\lambda} \ln \left[ \frac {\lambda^2
V_0}{2} \frac{1+w_B}{1-w_B} \right],
\end{equation}
for $-1 < w_B < 1$.
Substituting this into Eq. (\ref{rhophi}) to determine $\rho_\phi$, we see that this expression for $\phi$
corresponds to a background-dominated
universe with $\rho_B \gg \rho_\phi$ only in the limit where $\lambda \gg 1$; in this limit, Eq. (\ref{exppart}) is an attractor for all values
of $w_B$ in the range $-1 < w_B < 1$ \cite{CLW}.

\subsection{Exact solutions}

Now, however, we will see seek an exact solution for Eq. (\ref{phiexp}).  We make the change of variables
\begin{equation}
t = e^\tau
\end{equation}
and
\begin{equation}
\psi = \lambda \phi - 2 \tau
\end{equation}
and Eq. (\ref{phiexp}) becomes
\begin{equation}
\psi^{\prime \prime} + \left(\frac{2}{1+w_B} - 1 \right)(\psi^\prime + 2) - \lambda^2 V_0 e^{-\psi} = 0,
\end{equation}
where prime denotes the derivative with respect to $\tau$.
For the case of stiff matter with $w_B =1$ (only), the $\psi^\prime$ term vanishes, giving
\begin{equation}
\label{psieq}
\psi^{\prime \prime} - \lambda^2 V_0 e^{-\psi} = 0.
\end{equation}
This equation
can then be integrated to give the first integral
\begin{equation}
\label{psiexp}
\frac{1}{2} \psi^{\prime 2} + \lambda^2 V_0 e^{-\psi} = C,
\end{equation}
with constant $C$.  In terms of $\phi$ and $t$, this corresponds to
\begin{equation}
\frac{1}{2}(\lambda t \dot \phi - 2)^2 + \lambda^2 t^2 V_0 e^{-\lambda \phi} = C.
\end{equation}

Now note that Eq. (\ref{psieq}) can be integrated exactly, giving \cite{Polyanin}
\begin{equation}
\psi = 2 \ln \left[\sqrt{\frac{V_0}{2}}\frac{\lambda}{C_1}\cosh(C_1 \tau + C_2)\right],
\end{equation}
with $C_1$ and $C_2$ the constants of integration.
We can reexpress this exact solution in terms of $t$ and $\phi$ to obtain an exact solution to Eq. (\ref{phiexp})
with $w_B = 1$:
\begin{eqnarray}
\phi &=& \frac{2}{\lambda} \ln \left[\sqrt{\frac{V_0}{2}}\frac{\lambda t}{C_1} \cosh(C_1 \ln t + C_2)\right],\\
&=& \frac{2}{\lambda} \ln \left[\sqrt{\frac{V_0}{8}}\frac{\lambda}{C_1}(C_3 t^{C_1 + 1} + C_3^{-1} t^{1-C_1})\right],
\end{eqnarray}
where $C_3 = \exp(C_2)$.  A similar solution for $\lambda=1$ appears to have first been derived by Sajben \cite{Sajben}
for the equation describing the electron density near hot filaments in cylindrical coordinates.

\subsection{Applications}

In the simplest standard cosmological model, the universe undergoes periods of radiation and matter domination,
but the background energy density is never dominated by stiff matter with $w_B = 1$.  However, there has been increasing
interest in the possibility of such a period of stiff matter domination, also called ``kination" (since $w_B=1$ could
be driven by the kinetic energy of a scalar field).  The effects
of a kination/stiff-matter dominated era have been studied in relation to baryogenesis \cite{Joyce}, Big
Bang nucleosynthesis \cite{stiffBBN}, the relic
abundance of dark matter \cite{Salati,VG,Erickcek,Deramo}, and the propagation of gravitational radiation
(\cite{Gouttenoire} and references therein).  Hence, it is not unreasonable to also examine scalar field evolution in a background
with $w_B = 1$.

Unlike the power-law solutions, the exact solution for a scalar field with an exponential potential evolving
in a stiff-matter background is simpler and more
useful than the corresponding first integral.  To determine the values for
$C_1$ and $C_2$ corresponding to a given set of initial conditions, we will
take, for simplicity, $t=1$ to be our initial value of $t$, and $\phi = 0$
to be our initial value of $\phi$.  However, we will allow the initial
value of $\dot \phi$ to be a free parameter, $\dot \phi_i$.
Then the exact solution above gives the corresponding values for
$C_1$ and $C_2$, namely
\begin{eqnarray}
C_1 &=& \frac{\lambda}{2}\sqrt{2 V_0 + \left(\dot \phi_i- \frac{2}{\lambda} \right)^2},\\
C_2 &=& \sinh^{-1} \left[\left(\frac{1}{\sqrt{2V_0}}\right)\left(\dot \phi_i -
\frac{2}{\lambda}\right)\right].
\end{eqnarray}
Using our exact solution with Eq. (\ref{rhophi}), we can derive an expression
for the scalar field energy density,
\begin{equation}
\label{rhophiexp}
\rho_\phi = \frac{2}{\lambda^2}\frac{1}{t^2}[1 + C_1^2 + 2 C_1 \tanh(C_1 \ln t + C_2)],
\end{equation}
and for the equation of state parameter,
$w_\phi$,
\begin{equation}
\label{wphiexp}
w_\phi = 1 -  \frac{2 C_1^2}{(1+C_1^2)\cosh^2 (C_1 \ln t + C_2)
+ 2 C_1 \cosh (C_1 \ln t + C_2) \sinh (C_1 \ln t + C_2)}.
\end{equation}
It is clear from Eq. (\ref{wphiexp}) that the equation of
state parameter for the scalar field evolves asymptotically to $w_\phi \rightarrow 1$,
so that the scalar field energy density evolves toward that of
the background stiff matter.  This is illustrated
in Fig. 3, which shows the time evolution of $w_\phi$ from
Eq. (\ref{wphiexp}) for a variety of model parameters.
\begin{figure}[t]
\centerline{\epsfxsize=3.7truein\epsfbox{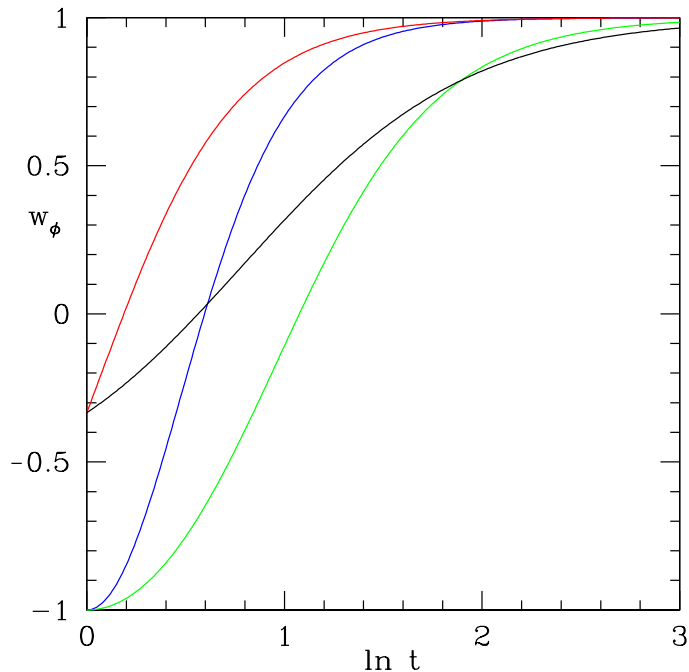}}
\caption{The evolution of the scalar field equation of
state parameter, $w_\phi$, as a function of $\ln t$ as given by the exact solution (Eq. \ref{wphiexp}) for the exponential
potential $V(\phi) = V_0 \exp(-\lambda \phi)$, with $\phi_i = 0$ at initial time
$t=1$, and $V_0 = 1$.  Curves correspond to $\dot \phi_i = 0$, $\lambda = 1$
(green), $\dot \phi_i = 0, \lambda = 2$ (blue), $\dot \phi_i = 1, \lambda = 1$
(black), and $\dot \phi_i =1$, $\lambda = 2$ (red), where $\dot \phi_i$ is
the initial value of $\dot \phi$ at $t=1$.}
\end{figure}
While the evolution to $w_\phi = 1$ is inevitable in this model, Eq. (\ref{wphiexp}) gives
insight into the rate at which this evolution occurs.  Larger values of $C_1$ correspond to a more
rapid increase in $w_\phi$, as is apparent in Fig. 3.  For fixed $\dot \phi_i$, this corresponds to larger values of both
$V_0$ and $\lambda$, while for fixed $V_0$ and $\lambda$, the minimum rate of increase of $w_\phi$ occurs
for $\dot \phi_i = 2/\lambda$.

In the late-time limit, $t \rightarrow \infty$, the scalar
field density given by Eq. (\ref{rhophiexp}) becomes
\begin{equation}
\rho_\phi = \frac{2}{\lambda^2}\frac{1}{t^2} (1+C_1)^2.
\end{equation}
In comparison, the density of the background stiff matter is $\rho_B = 1/(3t^2)$.
Thus, the background matter will dominate the expansion, and our
exact solution for the evolution of $\phi$ will remain valid arbitrarily long
when $2(1+C_1)^2/\lambda^2 < 1/3$.  When this equation is not satisfied,
the scalar field energy density will eventually come to dominate the expansion,
and our results will no longer be valid.  Beyond that point the evolution
will be given instead by the exact solution of Refs.
\cite{Russo,Andrianov,Piedipalumbo} for the exponential potential with a scalar field dominated
expansion.

\section{Discussion}

The results presented here can be generalized in several straightforward ways.  Adding a constant to any
potential leaves Eq. (\ref{phiback}) unchanged, so
we obtain the same solution $\phi(t)$, but with
a constant added to $\rho_\phi$.  Furthermore, translating any of the potentials by a constant
value of $\phi$ simply translates the corresponding solution by that same constant.

One set of models we have not chosen to examine are those in which the expansion is dominated by vacuum energy.  In this
case, we have $H = H_0$ (a constant), and instead of Eq. (\ref{phiback}), the evolution of $\phi$ is given by
\begin{equation}
\label{phivac}
\ddot{\phi} + 3H_0{\phi} + \frac{dV}{d{\phi}} = 0.
\end{equation} 
This would correspond, for example, to the evolution of a scalar field in a universe dominated by a cosmological constant.
However, from a mathematical point of view, it is clear that Eq. (\ref{phivac}) is qualitatively different from Eq.
(\ref{phiback}).  Hence, we have chosen to leave these models for future investigation.

In contrast to the well-explored particular attractor solutions, relatively less attention has been paid to exact solutions
of Eq. (\ref{phiback}).  We have found a variety of such solutions for both power-law and exponential potentials.  While we
cannot rule out exact solutions for more complicated potentials, we have likely exhausted the simplest exactly-solvable
cases.  Our exact solutions are applicable to a much narrower range of potentials than is the case for the
particular attractor solutions; unfortunately, one cannot pick and choose which potentials yield exact solutions.
However, since it is difficult to predict precisely which scalar field potentials will be of interest to future investigators,
it seems worthwhile to provide the general catalog given in this paper.

\begin{acknowledgments}

R.J.S. was supported in part by the Department of Energy (DE-SC0019207).

\end{acknowledgments}


\begin{thebibliography}{99}


\bibitem{Lyth}
D.H. Lyth and A.A. Riotto, Phys. Rept. {\bf 314}, 1 (1999).

\bibitem{Allahverdi}
R. Allahverdi, R. Brandenberger, F.-Y. Cyr-Racine,
and A. Mazumdar, Ann. Rev. Nucl. Part. Sci. {\bf 60}, 27 (2010).

\bibitem{RatraPeebles}
  B. Ratra and P.J.E. Peebles,
  Phys.\ Rev.\  D {\bf 37}, 3406 (1988).
  
\bibitem{Wetterich1}
C. Wetterich, Nucl. Phys. B {\bf 302}, 668 (1988).

\bibitem{Wetterich2}
C. Wetterich, Astron. Astrophys. {\bf 301}, 321 (1995).
  
\bibitem{Ferreira1} P.G. Ferreira and M. Joyce,
\prl {\bf 79}, 4740 (1997).

\bibitem{Ferreira2} P.G. Ferreira and M. Joyce,
\prd {\bf 58}, 023503 (1998).

\bibitem{CLW} E.J. Copeland, A.R. Liddle, and D. Wands,
\prd{\bf 57}, 4686 (1998).  
  

\bibitem{CaldwellDaveSteinhardt}
  R.R. Caldwell, R. Dave and P. J. Steinhardt,
  Phys.\ Rev.\ Lett.\  {\bf 80}, 1582 (1998).
  

\bibitem{LiddleScherrer}
  A.R. Liddle and R.J. Scherrer,
  Phys.\ Rev.\  D {\bf 59}, 023509 (1999).
  
  
\bibitem{SteinhardtWangZlatev}
  P.J. Steinhardt, L.M. Wang and I. Zlatev,
  Phys.\ Rev.\  D {\bf 59}, 123504 (1999).
  
\bibitem{Copeland1}
E.J. Copeland, M. Sami, and S. Tsujikawa, Int. J. Mod. Phys. D
{\bf 15}, 1753 (2006).

\bibitem{Bamba}
K. Bamba, S. Capozziello, S. Nojiri, and S.D. Odintsov,
Astrophys. Space Sci. {\bf 342}, 155 (2012).

\bibitem{Poulin}
V. Poulin, T.L. Smith, T. Karwal, and M. Kamionkowski,
Phys. Rev. Lett. {\bf 122}, 221301 (2019).

\bibitem{Agrawal}
P. Agrawal, F.-Y. Cyr-Racine, D. Pinner, and L. Randall,
arXiv:1904.01016.

\bibitem{Smith}
T.L. Smith, V. Poulin, and M.A. Amin,
\prd {\bf 101}, 063523 (2020).

\bibitem{Hill}
J.C. Hill, E. McDonough, M.W. Toomey, and S. Alexander,
\prd {\bf 102}, 043507 (2020).

\bibitem{Ivanov}
M.M. Ivanov, et al.,
Phys. Rev. D {\bf 102}, 103502 (2020).

\bibitem{Banerjee}
A. Banerjee, et al.,
Phys. Rev. D {\bf 103}, 081305 (2021).

\bibitem{Vagnozzi}
S. Vagnozzi,
\prd {\bf 104}, 063524 (2021).


\bibitem{Smith2}
T.L. Smith, et al.,
Phys. Rev. D {\bf 103}, 123542 (2021).

\bibitem{Hill2}
J.C. Hill, et al.,
arXiv:2109.04451.

\bibitem{Poulin3}
V. Poulin, T.L. Smith, and A. Bartlett,
\prd {\bf 104}, 123550 (2021).

\bibitem{SalopekBond}
D.S. Salopek and J.R. Bond, \prd {\bf 42}, 3936 (1990).

\bibitem{Russo}
J.G. Russo, Phys. Lett. B {\bf 600}, 185 (2004).

\bibitem{Elizalde}
E. Elizalde, S. Nojiri, and S.D. Odintsov,
\prd {\bf 70}, 043539 (2004).

\bibitem{Andrianov}
A.A. Andrianov, F. Cannata, and A. Yu. Kamenshchik,
JCAP {\bf 10}, 004 (2011).

\bibitem{Piedipalumbo}
E. Piedipalumbo, P. Scudellaro, G. Esposito, and C. Rubano,
Gen. Rel. Grav. {\bf 44}, 2611 (2012).

\bibitem{Harko}
T. Harko, F.S.N. Lobo, and M.K. Mak,
Eur. Phys. Jour. C {\bf 74}, 2784 (2014).

\bibitem{ScherrerSen}
R.J. Scherrer and A.A. Sen,
Phys.\ Rev.\  D {\bf 77}, 083515 (2008).

\bibitem{ds1}
S. Dutta and R.J. Scherrer,
\prd {\bf 78}, 123525 (2008).

\bibitem{Chiba}
T. Chiba, \prd {\bf 79}, 083517 (2009).
  

\bibitem{ds2}
  S. Dutta, E.N. Saridakis and R.J. Scherrer,
  \prd {\bf 79}, 103005 (2009).
  
\bibitem{Basilakos}
S. Basilakos, M. Tsamparlis, and A. Paliathanasis,
\prd {\bf 83}, 103512 (2011).

\bibitem{ACK}
A.A. Andrianov, F. Cannata, and A.Yu. Kamenshchik,
\prd {\bf 86}, 107303 (2012).
  
\bibitem{Linder}
E.V. Linder, Astropart. Phys. {\bf 24}, 391 (2005).

\bibitem{Sahlen}
M. Sahlen, A.R. Liddle, and D. Parkinson, \prd {\bf 72}, 083511 (2005).

\bibitem{masso}
E. Masso, F. Rota, and G. Zsembinszki,
\prd {\bf 72}, 084007 (2005). 

\bibitem{ds3}
S. Dutta and R.J. Scherrer,
\prd {\bf 78}, 083512 (2008).
  
\bibitem{Polyanin}
A.D. Polyanin and V.F. Zaitsev, {\it Exact Solutions
for Ordinary Differential Equations}, Chapman \& Hall/CRC (2003).

\bibitem{Binney}
J. Binney and S. Tremaine, {\it Galactic Dynamics},
Princeton University Press (1987).

\bibitem{Srivastava}
S. Srivastava,
\apj {\bf 136}, 680 (1962).

\bibitem{Sharma}
V.D. Sharma, Phys. Lett. A {\bf 60}, 381 (1977).

\bibitem{Mach}
P. Mach, J. Math. Phys. {\bf 53}, 062503 (2012).

\bibitem{Leach}
P.G.L. Leach, Jour. Math. Phys. {\bf 26}, 2510 (1985).

\bibitem{Greene}
P.B. Greene, L. Kofman, A.D. Linde, and A.A. Starobinsky, \prd {\bf 56},
6175 (1997).

\bibitem{Turner}
M.S. Turner, \prd {\bf 28}, 1243 (1983).

\bibitem{Sajben}
M. Sajben, Phys. Fluids {\bf 11}, 2501 (1968).

\bibitem{Joyce}
M. Joyce, \prd {\bf 55}, 1875 (1997).

\bibitem{stiffBBN}
S. Dutta and R.J. Scherrer,
\prd {\bf 82}, 083501 (2010).

\bibitem{Salati}
P. Salati, Phys. Lett. B {\bf 571}, 121 (2003).

\bibitem{VG}
L. Visinelli and P. Gondolo,
\prd {\bf 81}, 063508 (2010).

\bibitem{Erickcek}
K. Redmond and A.L. Erickcek,
\prd {\bf 96}, 043511 (2017).

\bibitem{Deramo}
F. D'Eramo, N. Fernandez, and S. Profumo,
JCAP {\bf 05}, 012 (2017).

\bibitem{Gouttenoire}
Y. Gouttenoire, G. Servant, and P. Simakachorn,
[arXiv:2111.0115].

\end{thebibliography}
\end{document}